\documentclass{article}
\title{Interpretation of Lorentz boosts in conformally deformed special relativity theory.}
\author{A.A. Deriglazov\footnote{alexei@ice.ufjf.br ~ On leave of
absence from Dept. Math. Phys., Tomsk Polytechnical University,
Tomsk, Russia.}}
\date{Dept. de Matematica, ICE, Universidade Federal de Juiz de Fora,\\
MG, Brazil.}
\begin{document}
\maketitle
\large
\begin{abstract}
Conformally deformed special relativity is mathematically consistent example of a theory with 
two observer independent scales. As compare with recent DSR proposals, it is formulated starting 
from the position space. In this work we propose interpretation of Lorentz boosts of the model as 
transformations among accelerated observers. We point further that the model can be considered as 
relativistic version of MOND program and thus may be interesting in context of dark matter 
problem.
\end{abstract}


\noindent
\section{Introduction}

In recent work [1] it has been proposed a model based on deformation of standard realization of the 
Lorentz group by means of special conformal transformation
$\Lambda_{\lambda}\equiv (U_{\lambda})^{-1}\Lambda (U_{\lambda})$, 
$U_{\lambda}: x^\mu\longrightarrow\frac{x^\mu+\delta^{\mu}{}_0\lambda x^2}{1-2\lambda x^0-\lambda^2x^2}$. 
The aim was to construct
consistent example of doubly special relativity (DSR) model [2,3] (formulated in position space 
starting from 
the beginning), i.e. a theory with underlying symmetry group being the Lorentz group, but with 
kinematical predictions different from that of special relativity. Mathematical consistency of the model 
has been discussed, in particular, it resolve the problem of total momentum for multi-particle 
system presented in others DSR proposals. In some sense, our solution of the problem is an opposite 
as compare with others 
DSR proposals (see [2-5] and references therein). The known DSR proposals are based on various 
non-linear 
realizations of the Lorentz group in space of conserved momentum, i.e. are formulated as a list of kinematical 
rules of a theory. Ordinary energy-momentum relation $(P^\mu)^2=-m^2$ is not 
invariant under the realization and is replaced by $[U(P^\mu)]^2=-m^2$.   
The central problem of DSR kinematics is consistent definition of total 
momentum for many-particle system. Actually, due to non linear form of the transformations, ordinary 
sum of momenta does not transform as the constituents. Different covariant rules proposed in the 
literature lead to hardly acceptable features [4]. For our model, the Lorentz group is realized 
non linearly both on position and on conjugated momentum spaces, but conserved four-momentum turns 
out to be 
different from the conjugated one. On space of conserved momentum one has ordinary realization 
of the Lorentz transformations, so the composition rule is ordinary sum, see [1]. On other hand, 
energy and momentum have nonstandard relation with measurable quantities (coordinates and velocities). 
It suggests that kinematical predictions of our model differ from that of  
special relativity theory, see also 
below\footnote{The difference among the canonical momentum and 
the conserved one implies an 
interesting situation in canonically quantized version of the theory. 
While the conjugated variables $(x, p)$ have the standard brackets, commutators 
of the coordinates $x^\mu$ with the energy and momentum $P^\mu$ are deformed. Thus the phase 
space $(x, P)$ is endowed with the noncommutative geometry (with the commutators $[x, P]$ 
and $[P, P]$ being deformed). In 
particular, the energy-momentum subspace turns out to be noncommutative.
The modified bracket $[x, P]$ suggests that the Planck's constant has slight 
dependence on $x$}. 

In the present work we propose interpretation of Lorentz boosts in the conformally deformed
special relativity (CDSR) model. In Sect. 2 we remind construction of deformed Lorentz realization 
as it was formulated in [1]. In Sect. 3 we demonstrate that the Lorentz boosts can be treated as 
transformations among mutually accelerated (in specific way) observers $O$ and $O'$. This 
interpretation is confirmed in Sect. 4, where we demonstrate that trajectory of $O'$ is particular 
solution of geodesic-line equation. It suggests that the model may be relevant for description of 
expanding universe, we speculate on this possibility in the Conclusion.     
 
\section{Conformally deformed Lorentz group realization}
  
In ordinary special relativity the requirement of invariance of the Minkowski 
interval: $ds^{'2}=ds^2$ immediately leads to the observer independent 
scale $|v^i|=c$. To construct a theory with one more scale, the invariance 
condition seems to be too restrictive. Actually, the most general transformations 
$x^\mu\longrightarrow x^{'\mu}(x^\nu)$ which preserve the interval are known to 
be Lorentz transformations in the standard realization [6] 
$x^{'\mu}=\Lambda^\mu{}_\nu x^\nu$, the latter does not admit one more invariant 
scale. So, one needs to relax the invariance condition keeping, as before, the 
speed of light invariant. It would be the case if $ds^2=0$ will imply 
$ds^{'2}=0$, which guarantees appearance of the invariant scale $c$ (in 
the case of linear relation $x^0=ct$). 

Thus, supposing existence of one more observer independent scale $R$, one assumes 
deformation of the invariance condition: $ds^{'2}=A(x,R)ds^2$, where 
$A \stackrel{R\rightarrow \infty}{\longrightarrow}1$. By construction, the maximum 
velocity remains the invariant scale of the formulation. In the limit 
$R\rightarrow \infty$ one obtains ordinary special relativity theory. 

Complete symmetry 
group for the case is the conformal group. It involves, 
in particular, special conformal transformations with the parameter $b^\mu$
\begin{eqnarray}\label{201}
U_b: x^\mu\longrightarrow\frac{1}{\Omega}(x^\mu+b^\mu x^2), \cr 
\Omega(x, b)\equiv 1+2b x+b^2x^2.\quad
\end{eqnarray}
Similarly to momentum DSR proposals [2, 3], let us deform the Lorentz group 
realization in accordance with the rule 
\begin{eqnarray}\label{202}
\begin{array}{l}
\Lambda_b\equiv (U_b)^{-1}\Lambda (U_b), \qquad \qquad \cr 
\Lambda_b: x^\mu\longrightarrow\frac{1}{G}\left[(\Lambda x)^\mu+
\left[(1-\Lambda)b\right]^\mu x^2\right], \quad \cr
G(x, b, \Lambda)\equiv 1-2b(1-\Lambda)x+2b(1-\Lambda)bx^2.
\end{array}
\end{eqnarray}
The above mentioned proportionality factor for the case is $A=G^{-2}$. 
The parameters $b^\mu$ can be 
further specified by the requirement that space rotations 
$\Lambda^\mu{}_\nu=( \Lambda^0{}_0 =1, ~  
\Lambda^0{}_i= \Lambda^i{}_0 =0, ~ \Lambda^i{}_j\equiv R^i{}_j, ~ 
R^T=R^{-1})$ are not deformed 
by $b^\mu$. Then the only choice is $b^\mu=(\lambda, 0, 0, 0)$, 
which gives final form of the deformed Lorentz group realization
\begin{eqnarray}\label{203}
\Lambda_\lambda: x^\mu\longrightarrow\frac{1}{G}\left[(\Lambda x)^\mu+
(\delta^\mu{}_0-\Lambda^\mu{}_0)\lambda x^2\right], \qquad \\
G(x, \lambda, \Lambda)\equiv 1+2\lambda(x^0-\Lambda^0{}_\mu x^\mu)
-2\lambda^2(1-\Lambda^0{}_0)x^2.
\end{eqnarray}
Our convention for the Minkowski metric is $\eta_{\mu\nu}=(-, +, +, +)$.
One confirms now emergence of one more observer independent scale:  
there is exist unique vector $x^\mu$ with zero component unaltered by 
the transformations (\ref{203}). Namely, from the condition 
$x^{'0}=x^0$ one has the only solution $x^\mu=(R\equiv -\frac{1}{\lambda}, 0, 0, 0)$ 
(the latter turns out to be the fixed vector). Thus all  
observers should agree to identify $R$ as the invariant scale. Let us point 
that the transformations (\ref{203}) are not equivalent to either the 
Fock-Lorentz realization [7], or to 
recent DSR proposals (the realizations lead to varying speed of light).

Inspection of transformation properties of quantities in our disposal allows one [1] 
to find invariant interval under the transformations (\ref{203}) 
\begin{eqnarray}\label{204}
ds^2=\frac{\eta_{\mu\nu}dx^\mu dx^\nu}{(1+2\lambda x^0-\lambda^2x^2)^2}
\equiv g_{\mu\nu}(x)dx^\mu dx^\nu,
\end{eqnarray}
On the domain where the metric is 
non degenerated, the corresponding four dimensional scalar curvature is zero, 
while three-dimensional space-like slice $x^0=0$ is curved space with  
constant curvature $R_{(3)}=-\frac{24}{R^2}$.

\section{Interpretation of Lorentz boosts in CDSR}

As it was mentioned above, space part of the deformed Lorentz transformations (\ref{203}) 
represents usual 
rotations, the latter are not deformed by the scale $\lambda$. Let us discuss the remaining part, 
corresponding to generators $M_{0i}$, i.e. Lorentz boosts of the model: 
$\Lambda^{\mu}{}_{\nu}=exp\left(2\omega^{0i}M_{0i}\right)^{\mu}{}_{\nu}$. 
Keeping $\omega^{01}\equiv\alpha\ne 0$ only, non zero matrix elements of 
$\Lambda^{\mu}{}_{\nu}$ are $\Lambda^{0}{}_{0}=\sinh\alpha, 
\Lambda^{0}{}_{1}=\Lambda^{1}{}_{0}=\cosh\alpha, \Lambda^{1}{}_{1}=\cosh\alpha, 
\Lambda^{2}{}_{2}=\Lambda^{3}{}_{3}=1$. Supposing further the standard relation $x^0=ct$, one 
obtains the following form of Eq.(\ref{203}), 
\begin{eqnarray}\label{301}
\begin{array}{l}
t'=\frac{1}{G}\left[\left(t+\lambda ct^2-\frac{\lambda}{c}x^2\right)\cosh\alpha+\frac{x}{c}\sinh\alpha+
\frac{\lambda}{c}\left(-c^2t^2+x^2\right)\right], \cr
x'=\frac{1}{G}\left[c\left(t+\lambda ct^2-\frac{\lambda}{c}x^2\right)\sinh\alpha+x\cosh\alpha\right], \cr
x'{}^2=0, \qquad \qquad x'{}^3=0, \cr
G\equiv 1+2\lambda\left[c\left(t+\lambda ct^2-
\frac{\lambda}{c}x^2\right)\left(1-\cosh\alpha\right)-x\sinh\alpha\right],
\end{array}
\end{eqnarray}
where it was taken $x^1\equiv x, ~ x^2=x^3=0$. 
Suppose that this expression corresponds to transformation low  
for observers $O$ and $O'$ in some state of motion in $x$-direction among themselves. 
Consistency of the picture 
requires that the transformation parameter $\alpha$ does not depend on $x, t$. We demonstrate that it 
can be achieved by the following choice of the state of motion
\begin{eqnarray}\label{302}
t+\lambda ct^2-\frac{\lambda}{c}\left(x(t)\right)^2=\frac{1}{V}x(t), \qquad V=const,
\end{eqnarray}
or, equivalently
\begin{eqnarray}\label{303}
\begin{array}{l}
x(t)=\frac{c}{2\lambda}\left[-\frac{1}{V}+\sqrt{\frac{1}{V^2}+
\frac{4\lambda}{c}\left(t+\lambda ct^2\right)}\right]= \cr
Vt+\lambda cV\left(1-\frac{V^2}{c^2}\right)t^2+O(\lambda^2), \cr
x(0)=0, \qquad \dot x(0)=V.
\end{array}
\end{eqnarray}
Choice of sign for the square root is dictated by the limit $x\rightarrow Vt$ as $\lambda\rightarrow 0$. 
Similarly to the special relativity case, the parameter 
$\alpha$ can be determined now as follows: suppose the systems $O, ~ O'$ coincide at $t=0$, with  
instantaneous relative velocity being $V$. Let $(t, x)$ be coordinates of some event, which happens 
at the origin of $O'$ at the moment $t$: $(t', x'=0)$. From second equation of the system (\ref{301}) one 
immediately obtains
\begin{eqnarray}\label{304}
0=\frac{1}{G}\left[\frac{c}{V}x(t)\sinh\alpha+x(t)\cosh\alpha\right], \Longrightarrow 
\tanh\alpha=-\frac{V}{c},
\end{eqnarray}
i.e. the standard SR relation. Substitution of this result into Eq.(\ref{301}) gives expression for the Lorentz 
boost in terms of (instantaneous) relative velocity among the observers
\begin{eqnarray}\label{305}
\begin{array}{l}
x'=G^{-1}\left(1-\frac{V^2}{c^2}\right)^{-\frac{1}{2}}\left(x-V\left(t+\lambda ct^2-
\frac{\lambda}{c}x^2\right)\right), \cr
t'=G^{-1}\left(1-\frac{V^2}{c^2}\right)^{-\frac{1}{2}}\left(t-\frac{V}{c^2}x+\lambda ct^2-
\frac{\lambda}{c}x^2\right)+\frac{\lambda}{G c}\left(-c^2t^2+x^2\right), \cr
G\equiv 1+2\lambda\left(c\left(t+\lambda ct^2-\frac{\lambda}{c}x^2\right)
\left(1-\left(1-\frac{V^2}{c^2}\right)^{-\frac{1}{2}}\right)+\right. \cr
\left.\left(1-\frac{V^2}{c^2}\right)^{-\frac{1}{2}}\frac{V}{c}x\right).
\end{array}
\end{eqnarray}
One concludes that Lorentz boosts of the model describe 
transformation among accelerated (according to Eq.(\ref{303})) observers. In the next section we demonstrate 
that it is reasonable interpretation. Namely, 
analysis of geodesic motion of a particle in the model shows that inertial motion is not the geodesic one, 
i.e. mutually inertial observers are not free.    
In contrast, Eq.(\ref{303}) turns out to be solution of geodesic equations of motion. 

As it should be, the transformations obtained coincide with the Lorentz boosts in the limit 
$\lambda\rightarrow 0$. Note also that Eq.(\ref{303}) is
consistent with the maximum signal velocity. Actually 
\begin{eqnarray}\label{30}
v=\frac{dx}{dt}=\frac{1+2\lambda ct}{\sqrt{\frac{1}{V^2}+
\frac{4\lambda}{c}\left(t+\lambda ct^2\right)}}, 
\end{eqnarray}
from which it follows $v\rightarrow c$ as $t\rightarrow\infty$. Then an observer with initial velocity 
$V<c$ will has velocity $v$ less than $c$ in the future, as it should be. Let us point, that this 
picture suggests interpretation of singularity presented in Eqs.(\ref{204}), (\ref{301}) in terms of 
event horizon of an observer. 

\section{Inertial observers are replaced by accelerated ones in CDSR}

The invariant interval (\ref{204}) suggests the following  
action for a particle motion [1] 
\begin{eqnarray}\label{401}
S=\frac{1}{2}\int d\tau\left[\frac{\eta_{\mu\nu}\dot x^\mu\dot x^\nu}
{e(1+2\lambda x^0-\lambda^2x^2)^2}-em^2\right]. 
\end{eqnarray}
It is invariant under the global symmetry (\ref{203}), under the "translations": 
$x^{'\mu}=(SC_\lambda)^{-1}e^{a^{.}\partial}(SC_\lambda)x^\mu$ with the 
parameters $a^\mu$, 
as well as under the reparametrizations $\tau\longrightarrow\tau^{'}(\tau), ~ 
x^{'\mu}(\tau^{'})=x^\mu(\tau), ~ 
e^{'}(\tau^{'})=\frac{\partial\tau}{\partial\tau^{'}}e(\tau)$.
Hamiltonian formulation of the theory has been described in some details in [1].
Dynamics is governed by the following equations of motion and constraint:  
\begin{eqnarray}\label{402}
\dot x^\mu=\tilde\Omega^2p^\mu, \quad 
\dot p^\mu=-\frac{2m^2}{\tilde\Omega}(\delta^\mu{}_{0}\lambda+\lambda^2x^\mu),
\end{eqnarray}
\begin{eqnarray}\label{403}
p^2=-\widetilde\Omega{}^{-2}m^2.
\end{eqnarray}
where $\widetilde\Omega\equiv(1+2\lambda x^0-\lambda^2x^2)$, ~ 
$p$ represents conjugated momentum for $x$, and the standard gauge $e=1$ for the constraint 
$p_e=0$ has been chosen. 

Let us find equations of motion for the physical variables $x^i(t)$, where $t=\frac{x^0}{c}$. To this end 
one needs to fix a gauge for the first class constraint (\ref{403}). Note that the standard gauge 
$x^0=p^0\tau$ is not covariant, since $x^0$ and $p^0$ have different transformation low. The 
covariant gauge turns out to be (see [1] for interpretation of this expression) 
\begin{eqnarray}\label{404}
\begin{array}{l}
x^0=\widetilde\Omega\left[\widetilde\Omega\left(p^0-2\lambda(xp)\right)-
2\left(x^0-\lambda x^2\right)\left(\lambda p^0-\lambda^2(xp)\right)\right]\tau+ \cr 
\lambda(xx)=p^0\tau+ O(\lambda).
\end{array}
\end{eqnarray}
It can be shown that the gauge is consistent with Eq.(\ref{402}). In the gauge chosen equations of 
motion for $x^i(x^0), ~ p^i(x^0)$ can be written in the form
\begin{eqnarray}\label{405}
\begin{array}{l}
\dot x^i=\frac{\widetilde\Omega p^i}{m}
\left(1+\left(\frac{\widetilde\Omega p^j}{m}\right)^2\right)^{-\frac{1}{2}}, \cr
\dot p^i=-2m\lambda^2\widetilde\Omega^{-2}x^i
\left(1+\left(\frac{\widetilde\Omega p^j}{m}\right)^2\right)^{-\frac{1}{2}},
\end{array}
\end{eqnarray}
or, equivalently
\begin{eqnarray}\label{406}
\begin{array}{l}
m\widetilde\Omega^{-1}\left(1-(\dot x^j)^2\right)^{-\frac{1}{2}}\dot x^i=p^i, \cr
\dot p^i=-2m\lambda^2\widetilde\Omega^{-2}\left(1-(\dot x^j)^2\right)^{\frac{1}{2}}x^i.
\end{array}
\end{eqnarray}
It implies the following equations for $x^i(t)$
\begin{eqnarray}\label{407}
\begin{array}{l}
\frac{d^2x^i}{dt^2}=2\lambda c\widetilde\Omega^{-1}\left(1-\left(\frac{\dot x^j}{c}\right)^2\right)
\left(\left(1+\lambda ct\right)\dot x^i-\lambda cx^i\right). \cr
\widetilde\Omega\equiv 1+2\lambda ct+ \lambda^2c^2t^2-\lambda^2(x^i)^2.
\end{array}
\end{eqnarray}
In the first order on $\lambda$ one has
\begin{eqnarray}\label{12}
\ddot x^i=2\lambda c\left(1-\left(\frac{\dot x^j}{c}\right)^2\right)\dot x^i+O(\lambda^2).
\end{eqnarray} 
One notes that $x^i=d^it+b^i$ is not a solution of the equations. In contrast, it can be verified by 
direct computation that the trajectory (\ref{303}) of observer $O'$ obeys the equation (\ref{407}) 
and represents example of geodesic line of the model. Thus the Lorentz boost 
(\ref{305}) describe transformation among geodesically moving observers, the latter replace  
inertially moving observers of special relativity. 

\section{Conclusion}

As it was discussed, CDSR model can be considered as special relativity theory of accelerated 
according to Eq.(\ref{303}) observers. This expression turns out to be solution of geodesic-line 
equation of the theory (\ref{407}), the parameter $\lambda$ then has interpretation as a rate of 
expansion, and can be identified\footnote{It is not in contradiction with discussion in the end of 
Sect. 2, since our analysis is the local one (metric (\ref{203}) is singular).}  
with the Hubble constant $\lambda c\sim H_0$. It suggests 
modification of ordem $\lambda$ of Newtonian dynamics in average-velocities region in accordance with 
Eq.(\ref{303}): $x(t)=Vt+H_0V\left(1-\frac{V^2}{c^2}\right)t^2=Vt+a_0t^2$. 
Departure from Newton laws 
occurs in the limit of small accelerations $a\sim a_0$ of test particles. Let us point that such 
a kind modification of 
non relativistic dynamics (MOND program) has been quite successfull in explaining of rotational curves 
of galaxies [8] without introducing of dark matter. So, our suggestion is that CDSR model may represent 
relativistic basis for the MOND program.    

\section{Acknowledgments}
Author would like to thank the Brazilian foundations CNPq and FAPEMIG 
for financial support.

\end{document}